\documentclass[10pt,letterpaper]{article}
\usepackage{opex3}
\usepackage{color}

\begin{document}

\title{Interaction of laser-cooled $^{87}$Rb atoms with higher order modes of an optical nanofibre}

\author{Ravi Kumar$^{1,2}$, Vandna Gokhroo$^1$, Kieran Deasy$^1$, Aili Maimaiti$^{1,2}$, Mary C. Frawley$^{1,2}$, Ciar{\'a}n Phelan$^1$ and S{\'i}le Nic Chormaic$^{1,*}$}
\address{$^1$Light-Matter Interactions Unit, OIST Graduate University, Onna-son, Okinawa 904-0495, Japan \\
$^2$Physics Department, University College Cork, Cork, Ireland}

\email{$^*$sile.nicchormaic@oist.jp} 



\begin{abstract*}
Optical nanofibres are used to confine light to subwavelength regions and are very promising tools for the development of optical fibre-based quantum networks using cold, neutral atoms. To date, experimental studies on atoms near nanofibres have focussed on fundamental fibre mode interactions. In this work, we demonstrate the integration of a few-mode optical nanofibre into a magneto-optical trap for $^{87}$Rb atoms. The nanofibre, with a waist diameter of $\sim$700 nm, supports both the fundamental and first group of higher order modes and is used for atomic fluorescence and absorption studies. In general, light propagating in  higher order fibre modes has a greater evanescent field extension around the waist in comparison with the fundamental mode. By exploiting this behaviour, we demonstrate that the detected signal of fluorescent photons emitted from a cloud of cold atoms centred at the nanofibre waist is larger ($\sim$6 times) when higher order guided modes are considered as compared to the fundamental mode. Absorption of on-resonance, higher order mode probe light by the laser-cooled atoms is also observed.  These advances should facilitate the realisation of atom trapping schemes based on higher order mode interference.
\end{abstract*}

\section{Introduction}
Subwavelength diameter optical fibres, commonly known as ``optical nanofibres (ONFs)'', are proving to be of immense value for both fundamental and applied research with many different systems being investigated, such as cold atom manipulation and trapping \cite{Morr2009, Vetsch2010, Goban2012, Hakuta2012, Schneeweiss2014, Daly2014}, colloidal particle manipulation \cite{Brambilla2007, Lei2011, Frawley2014}, and sensing \cite{Tong2011, Wang2011}.  ONFs have a large evanescent field extension outside their waist region, making  them ideal for light-matter interactions studies. The integration of ONFs into atomic systems has been a focus of ever increasing research interest in recent years \cite{Morrissey2013}. Earlier experiments, such as those reported in \cite{Morr2009, Vetsch2010, Goban2012, Hakuta2012, Schneeweiss2014}, focussed on (i) the interaction of the light guided in the fundamental fibre mode, HE$_{11}$, with atoms, or (ii) excitation of the HE$_{11}$ mode through fluorescence coupling from resonantly excited atoms.  While the latter experimental technique provides a means of characterizing the atoms near the surface of the optical nanofibre \cite{Morr2009, Nayak2008, Das2010, Russell2013}, the former permits scenarios whereby atoms can be trapped around the optical nanofibre when far-detuned light is coupled into it \cite{Vetsch2010, Goban2012, Schneeweiss2014}. Such ONFs can be termed as \textit{single-mode ONFs (SM-ONFs)} since only the fundamental guided mode is supported \cite{Kien2005}. The advantage of such a system is that, in addition to atom trapping applications, the nanofibre provides an optical interface that may be exploited for quantum communication using ensembles of laser-cooled atoms \cite{Daly2014}.

\begin{figure}[htbp]
\centering\includegraphics[width=14cm]{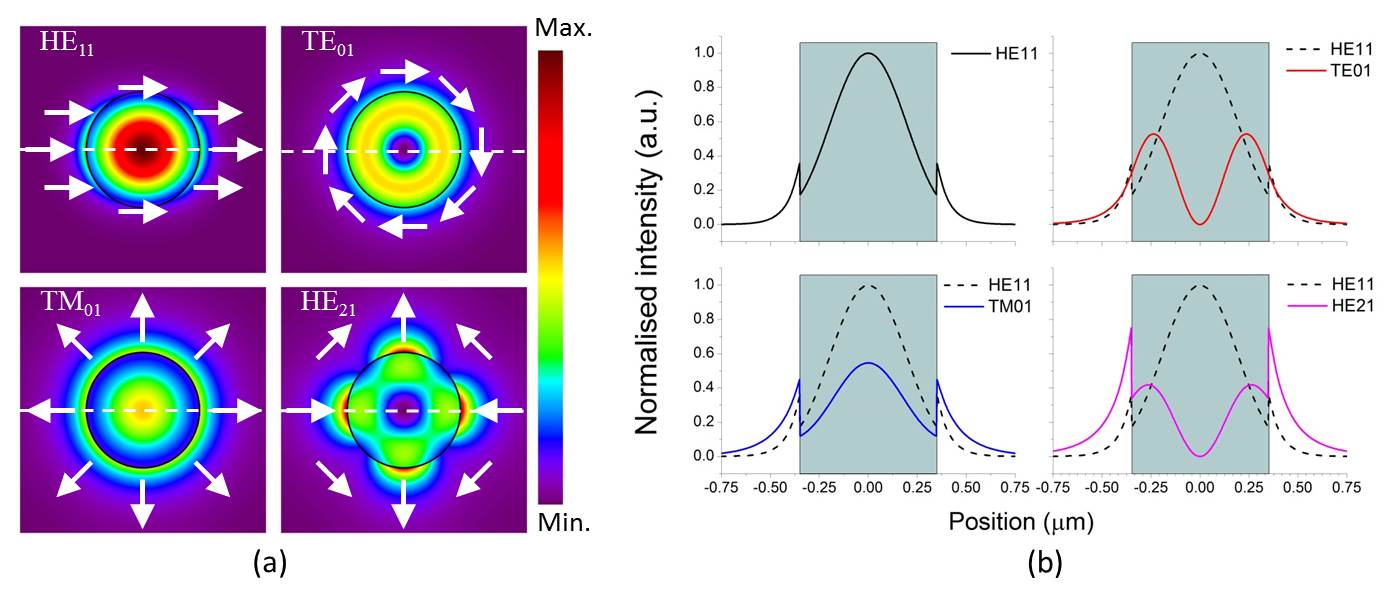}
\caption{ \label{fig:HM beam profile} (a) Intensity distribution of the first  four fibre modes (HE$_{11}$, TE$_{01}$, TM$_{01}$, and HE$_{21}$) of a nanofibre (radius 350 nm) for 780 nm light. Refractive indices of the nanofibre material and the surrounding medium are taken as 1.456 and 1 respectively. Black circles denote the nanofibre boundary. Arrows denote the polarization of the light fields. The amount of power in each mode is identical and the four plots are normalized to the maximum intensity in the HE$_{11}$ mode. (b) Intensity profile along the white dashed line in figure (a) with the zero position at the fibre axis. The shaded region denotes the fibre.   The intensities are normalized to the maximum intensity in the HE$_{11}$  mode. The profile of the HE$_{11}$ mode is plotted along with the other modes for ease of comparison.}
\end{figure}

Other atom trapping geometries based on the use of higher order modes (HOMs) in a nanofibre have been proposed, permitting greater flexibility of atom position relative to the fibre and relative to other trapped atoms \cite{ ref17, Fu2008, Sague2008}. These schemes have the benefit of allowing for selective mode interference by adjusting the trapping parameters according to experimental requirements. Efficient guiding of HOMs in ONFs was a major technical challenge until our recent reporting of low-loss mode propagation in nanofibres fabricated from 80 $\mu$m diameter silica fibre \cite{Frawley2012}.  Similar work using 50 $\mu$m fibre has since been reported \cite{Ravets2013a}. We term such fibres \textit{few-mode ONFs (FM-ONFs)} to distinguish them from the more conventional SM-ONFs.  The crucial step in advancing  such experiments was the result of a thorough study of the ideal parameters for fibre tapering, which revealed that reducing the fibre cladding-to-core diameter ratio relaxes the adiabatic criteria, thereby promoting efficient guiding of higher order modes \cite{Petcu-Colan2011}. The subsequent experimental achievements opened up a plethora of potential atom trapping scenarios, taking advantage of the few mode behaviour of the nanofibre. If one considers the field distribution of the first four true nanofibre modes (illustrated in Fig. \ref{fig:HM beam profile}), it is evident that the evanescent field for TE$_{01}$, TM$_{01}$ and HE$_{21}$ extends further into the surrounding medium than for  HE$_{11}$. For a nanofibre of diameter 700 nm, as used for the experiments reported in this work, the fraction of light outside the nanofibre is higher for each of the TE$_{01}$, TM$_{01}$ and HE$_{21}$ modes as compared with HE$_{11}$. This phenomenon results in the evanescent light field for the higher order modes interacting with more atoms in the surrounding cloud than when studies are limited to the fundamental mode.

In this work, we present the first demonstration of a few-mode ONF integrated into a cloud of laser-cooled $^{87}$Rb atoms.  The higher order modes in the ONF were maintained despite the integration of the nanofibre into the ultra high vacuum system. We use the FM-ONF for two different studies to contrast the difference between the higher and fundamental modes: (i) we consider the count rate of atomic fluorescence coupled into the fibre through the nanofibre waist for both fundamental and higher modes, and (ii) we measure the atomic absorption of an on-resonance probe guided in the fundamental or in the higher modes.  We show that the count rates for photons coupled into the nanofibre are larger when we consider higher order guided modes  than for the fundamental mode. This is in qualitative agreement with earlier theoretical predictions  \cite{Masalov2013, Masalov2014}. Note that the theory is based on single atom coupling to the nanofibre, whereas our system involves multiple atoms. Our studies also show that more atoms absorb light from the evanescent field when higher order modes are used, leading to better absorption signals (measured as a percentage of the probe light sent through the nanofibre). Aside from the impact this work will have on trapping schemes for cold atoms, it is also expected to drive progress in several other areas of research including optical communication, neutral-atom based quantum networks, particle manipulation, and sensing.

This paper is organised as follows. Section 2 describes the fabrication of the FM-ONF and the experimental details of the magneto-optical trap for $^{87}$Rb. Section 3 presents the methods and results we obtained for three different experiments: (i) coupling of light (from MOT beams) to the nanofibre in the absence of an atom cloud, (ii) coupling of light to the nanofibre in the presence of an atom cloud, and (iii) absorption of a nanofibre-guided, probe beam by the laser-cooled atoms. The conclusion is presented in Section 4.

\section{Experiment}
\subsection{Higher order mode optical nanofibre}

Most  published work on ONFs and atoms has focussed on  SM-ONFs.  Here, we used a FM-ONF which was  fabricated from an 80 $\mu$m  diameter, commercial, few-mode fibre for 780 nm (SM1250G80, Thorlabs) using the heat-and-pull technique, as previously described in \cite{Frawley2012}. We use 80 $\mu$m fibre in these experiments since it is easy to integrate with other fibre components using standard splicing techniques.  The fibre pulling rig used was a hydrogen-oxygen flame-brushed system \cite{Ward2014}. A 2 cm length of protective jacket was removed from the fibre and the two pigtails were clamped on translation stages on the  pulling rig. The flame heated the fibre near to its phase transition temperature ($1550\,^{\circ}\mathrm{C}$). The flame was brushed along the fibre for a particular set length (the ``hot zone'') while the fibre was simultaneously pulled by two translational stages moving in  opposite directions with a constant speed (the ``pulling speed''). The obtained taper profile was very close to  exponential in the taper regions. Higher order mode propagation is extremely sensitive to the taper angle, which itself depends on the pulling speed and hot zone. A longer hot zone facilitates shallower exponential tapers, leading to a higher transmission, but this also elongates the taper length. If the nanofibre is too long, it is not suitable for integration into our cold atom setup. For these reasons, and in order to achieve reasonably high transmission of the higher order modes through the nanofibre, an optimal hot zone of 7.7 mm and tapering speed of 0.125 mm/s were used. A photodetector and a charged coupled device (CCD) camera were placed at the output of the fibre to monitor the transmission and mode profile during the tapering process.

\begin{figure}[htbp]
\centering
\includegraphics[width=11 cm]{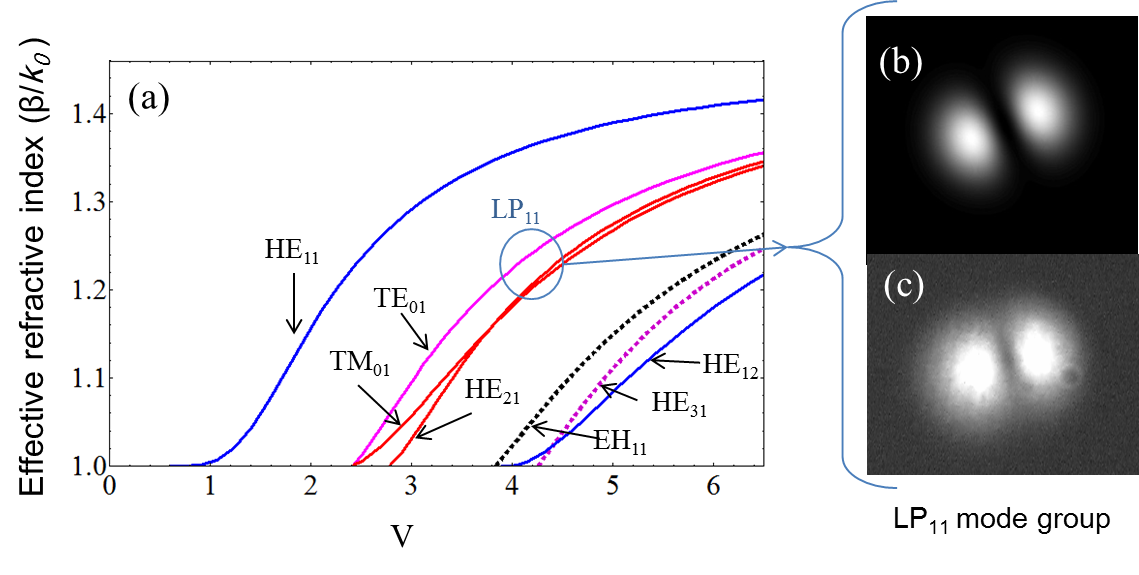}
     \caption { \label{fig:v_number} (a) V-number plot for different modes of the nanofibre with the V-parameter on the x-axis. The vertical line indicates the corresponding V-number for the nanofibre size used in the experiments;  (b) Simulated  profile for the LP$_{11}$ mode exiting from a nanofibre pigtail; (c) Observed experimental profile for (b). }
\end{figure}

 Light propagating in the linearly polarised, LP$_{11}$, approximate mode in the non-tapered fibre segments must be described by the the TE$_{01}$, TM$_{01}$ and HE$_{21}$ true modes in the nanofibre region, and LP$_{01}$ has its equivalence as the HE$_{11}$ mode.  For ease of notation, in the following discussions, we refer to the LP$_{11}$ approximate mode for indicating the family of true modes collectively, while recognising that the true modes provide us with the correct solutions.   Solving Maxwell's equations for an optical fibre \cite{Yariv1991} yields a waveguide mode parameter, called the V-number, from which the number of modes supported by the fibre can be obtained. The V-number is given as
\begin{equation}
V = \frac{2 \pi a}{\lambda} \sqrt{n_{core}^2 - n_{clad}^2} \,\,,
\end{equation}
where $a$ is the radius of the core, $\lambda$ is the wavelength of the light propagating in the fibre, and n$_{core}$ and n$_{clad}$ are the refractive indices of the core and the cladding, respectively. The V-number for the specific fibre used in our experiments (when untapered) is 4.3, implying that it can support four linearly polarised mode groups, LP$_{01}$, LP$_{11}$, LP$_{21}$, and LP$_{02}$, for 780 nm light.  When a nanofibre is fabricated, the cladding of the untapered fibre becomes the core for the nanofibre and the surrounding medium (e.g. air or vacuum) becomes the cladding. The V-number plot is shown in Fig. \ref{fig:v_number}. The mode cutoff diameter is around 660 nm for the degenerate HE$_{21}$ true modes and around 580 nm for both the TE$_{01}$ and TM$_{01}$ true modes.

\begin{figure}[htbp]
\centering \includegraphics[width=13 cm]{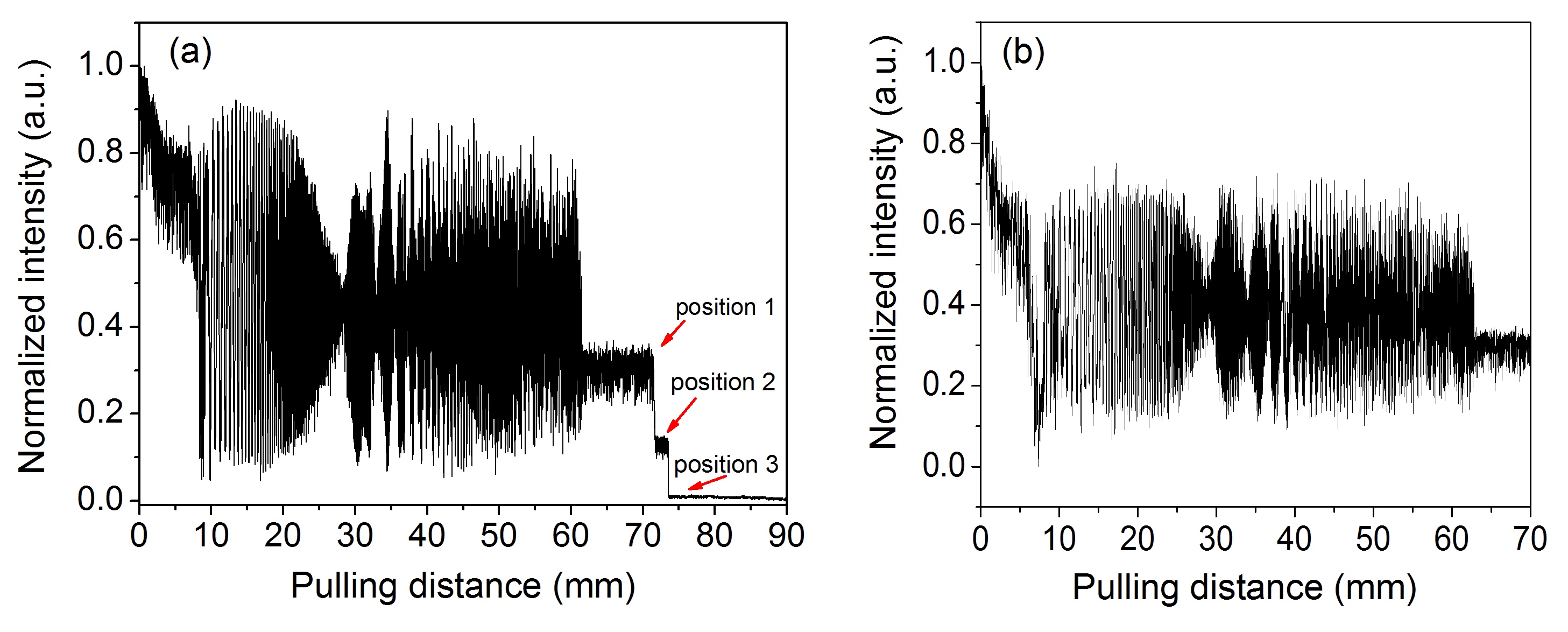}
     \caption{\label{fig:tranmission plot}
Observed transmission  for  light coupled to the LP$_{11}$ mode of the fibre during tapering: (a) for a pull length of 90 mm  to cut off all the higher modes and (b) for a 70 mm pulling length to ensure that the nanofibre supports only the TE$_{01}$, TM$_{01}$, HE$_{21}$, and HE$_{11}$ modes. In (a) Position 1 is the cutoff point for the coupled HE$_{21}$ modes, Position 2 is the cutoff  for the TE$_{01}$ and TM$_{01}$ modes and Position 3 is the portion of the residual fundamental mode guided by the nanofibre.}

\end{figure}

In our experiments, the LP$_{11}$ mode was excited by injecting a Laguerre-Gaussian (LG) beam into the untapered fibre pigtails \cite{Volpe2004}. A liquid-crystal-on-silicon spatial light modulator (Holoeye Pluto SLM) was used to create the LG beam \cite{Matsumoto2008}. A computer-generated vortex hologram was applied to the SLM and a vertically polarised, 780 nm laser beam was launched on to the SLM surface; the reflected beam formed a doughnut shape at the far field. This LG$_{01}$ free space beam was coupled to the fibre to excite the LP$_{11}$ mode which yielded a two-lobed beam profile at the fibre output, as expected. Along with the excitation of the LP$_{11}$ mode, there was a small percentage of  residual fundamental mode coupled into the fibre. This was mainly due to the purity of the generated LG beam and the efficiency of the fibre coupling. The level of the impurity was estimated by sending an LG$_{01}$ beam into the fibre during the tapering process and observing its transmission as a function of pulling distance (see Fig. \ref{fig:tranmission plot}(a)). The mode cutoff pull length for the higher modes (HE$_{21}$, TE$_{01}$, TM$_{01}$) and the fundamental mode can be defined from Fig. \ref{fig:tranmission plot}(a). Deducting the percentage of the remaining fundamental mode (Position 3) from the total intensity (Position 1), we estimated that the LP$_{11}$ mode was excited in the few-mode fibre with a purity of 95\%.

 First, a nanofibre was fabricated using a 90 mm pulling length in order to obtain a complete transmission profile including mode cutoffs. The transmission graph (Fig. \ref{fig:tranmission plot}(a)) suggested that the pulling distance should be between 62 and 71 mm to obtain a nanofibre supporting the LP$_{11}$ family of modes and the fundamental mode (Ref. \cite{Frawley2012} gives the details of the process). Since a smaller diameter fibre yields greater extension of the evanescent field into the surrounding medium, it is preferable to use a longer pulling length. For all other experiments reported here, we used a 70 mm pulling length to fabricate  the nanofibre.  This corresponded to $\sim$700 nm waist diameter determined from the calibration data of our pulling rig. Monitoring the mode profile until the end of the pulling process ensured that the prepared  fibre still supported the LP$_{11}$ mode. The nanofibre had 32\% transmission for the LP$_{11}$ mode group (see Fig. \ref{fig:tranmission plot}). The rest of the power was lost as it coupled to the cladding modes in the taper regions \cite{Ravets2013}. This fibre was highly adiabatic for the fundamental mode with 92\% transmission. Using these values of total transmissions (and assuming both sides of the taper were equivalent as verified by measurement), we estimate that 96\% of the input power was transmitted to the nanofibre waist for the fundamental mode and 56\% for the LP$_{11}$ mode group.

The nanofibre was glued to a U-shaped mount and installed vertically in an octagonal vacuum chamber used for the magneto-optical trap (MOT). The six cooling beams intersected at the nanofibre waist, four at 45$^{\circ}$ and two  at 90$^{\circ}$ to the nanofibre axis.  The fibre pigtails at either end of the nanofibre were passed through Teflon ferrules (with hand-drilled holes of 0.25 mm diameter) located on the top and the bottom flanges of the chamber. The ferrules were rendered vacuum tight by compressing Swagelok connectors \cite{Abraham1998}. A relatively simple experiment was carried out in order to test whether the distribution of power between the fundamental and the LP$_{11}$ modes was affected by pressure on the tapered fibre via the Teflon ferrule. As described earlier, an LG$_{01}$ mode was coupled to the fibre and the output was measured with a CCD camera.  We observed minimal mode mixing as the ferrule was tightened. Note that it is crucial to keep the nanofibre straight to ensure minimal bending loss and distortion of the mode profiles of the light passing through it.

\subsection{Magneto-optical trapping of atoms}

 $^{87}$Rb atoms were cooled and trapped using a standard magneto-optical trapping technique (for details see \cite{ Russell2013} though it was for  $^{85}$Rb). Base pressure in the vacuum chamber was 2$\times$10$^-$$^9$ mbar and, when a current of 5 A was passed through a Rb dispenser, the pressure rose to 4$\times$10$^-$$^9$ mbar and remained stable during the experiments.

 \begin{figure}[htbp]
\centering \includegraphics[width=13.5 cm]{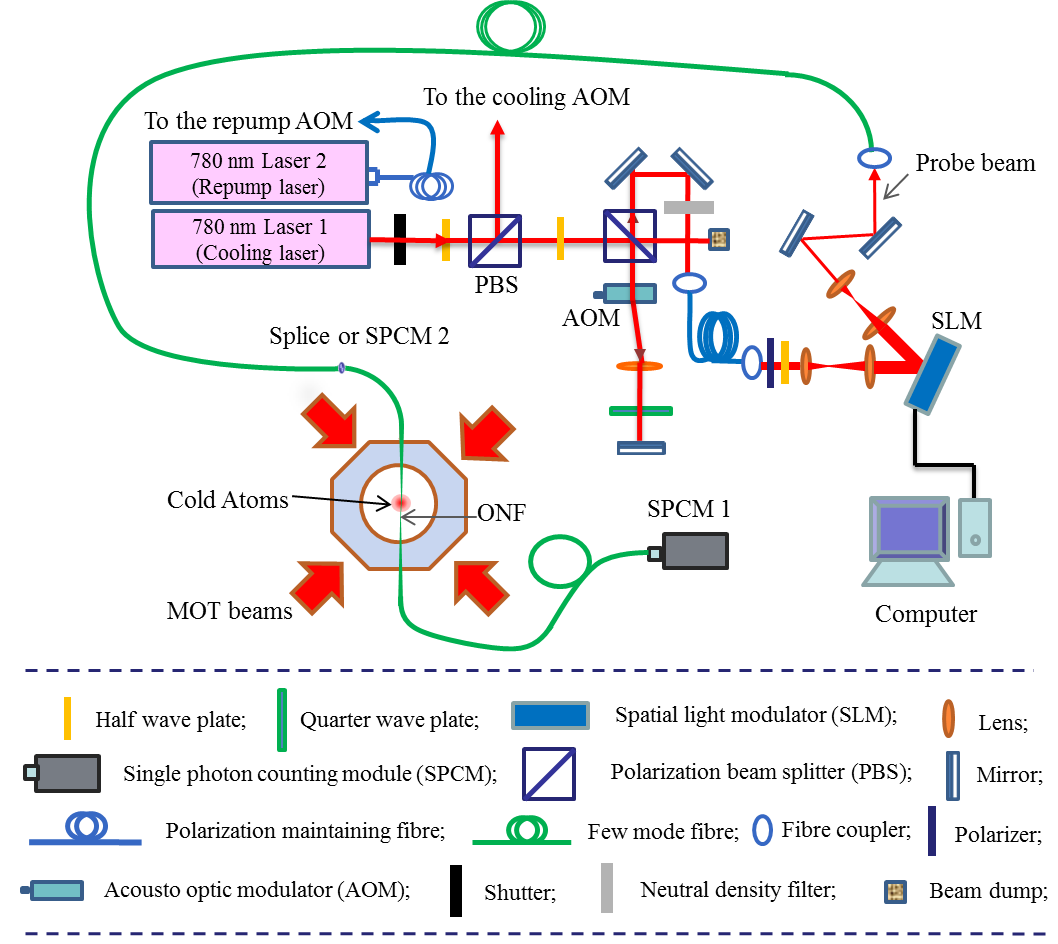}
     \caption{\label{fig:Schematic of the experimental setup} Schematic of the experimental setup. For absorption experiments one pigtail of the FM-ONF was spliced to a length of fibre (the same kind which was used to prepare the FM-ONF) transporting the probe light and another pigtail was connected to an SPCM.  For fluorescence measurements, the pigtails were connected to different devices according to the experiment being conducted (see Section 3.2). }
\end{figure}

A 780 nm laser was locked to the 5 $ ^2S_{1/2} F_g=2\rightarrow$ 5 $^2P_{3/2} F_e=(2,3)_{co}$ crossover peak of $^{87}$Rb for the cooling beam. The frequency was further shifted by an acousto-optical modulator (AOM) in a double-pass configuration so that it was 14 MHz red-detuned from the cooling transition, 5 $^2S_{1/2} F_g=2\rightarrow$5  $^2P_{3/2} F_e=3$. The cooling beam was split into four beams, two of which were retro-reflected to get three pairs of $\sigma^+$ and $\sigma^-$ beams for the MOT. Another laser, used as the repump, was locked to the 5 $^2S_{1/2} F_g=1\rightarrow$ 5 $^2P_{3/2} F_e=(0,2)_{co}$ peak and shifted to the repump transition, 5 $^2S_{1/2} F_g=1 \rightarrow$ 5 $^2P_{3/2} F_e=2$, using an AOM. The repump was overlapped with one of the cooling beams using a beam splitter. The magnetic field for the MOT was created by a pair of  coils carrying equal currents of 3.5 A in opposite directions to generate a field gradient of 10 G/cm at the centre of the vacuum chamber. Each cooling beam had an intensity of 6 mW/cm$^2$ and a diameter of 18 mm. The cold atoms were trapped around the waist of the FM-ONF. A compensation coil was used to generate a small magnetic field in the transverse direction to the MOT coils' axis in order to optimise the overlap between the centre of the atom cloud and the nanofibre. The diameter of  the cloud was $\sim$1 mm and there were $\sim$3$\times$10$^{6}$ atoms in the trap. The temperature of the cloud was measured to be $\sim$150 $\mu$K by taking a series of fluorescence images of the cloud for different expansion times.

\section{Measurements and results}
\subsection{Coupling of MOT beams to the nanofibre}
\begin{figure}[htbp]
\centering \includegraphics[width= 11 cm]{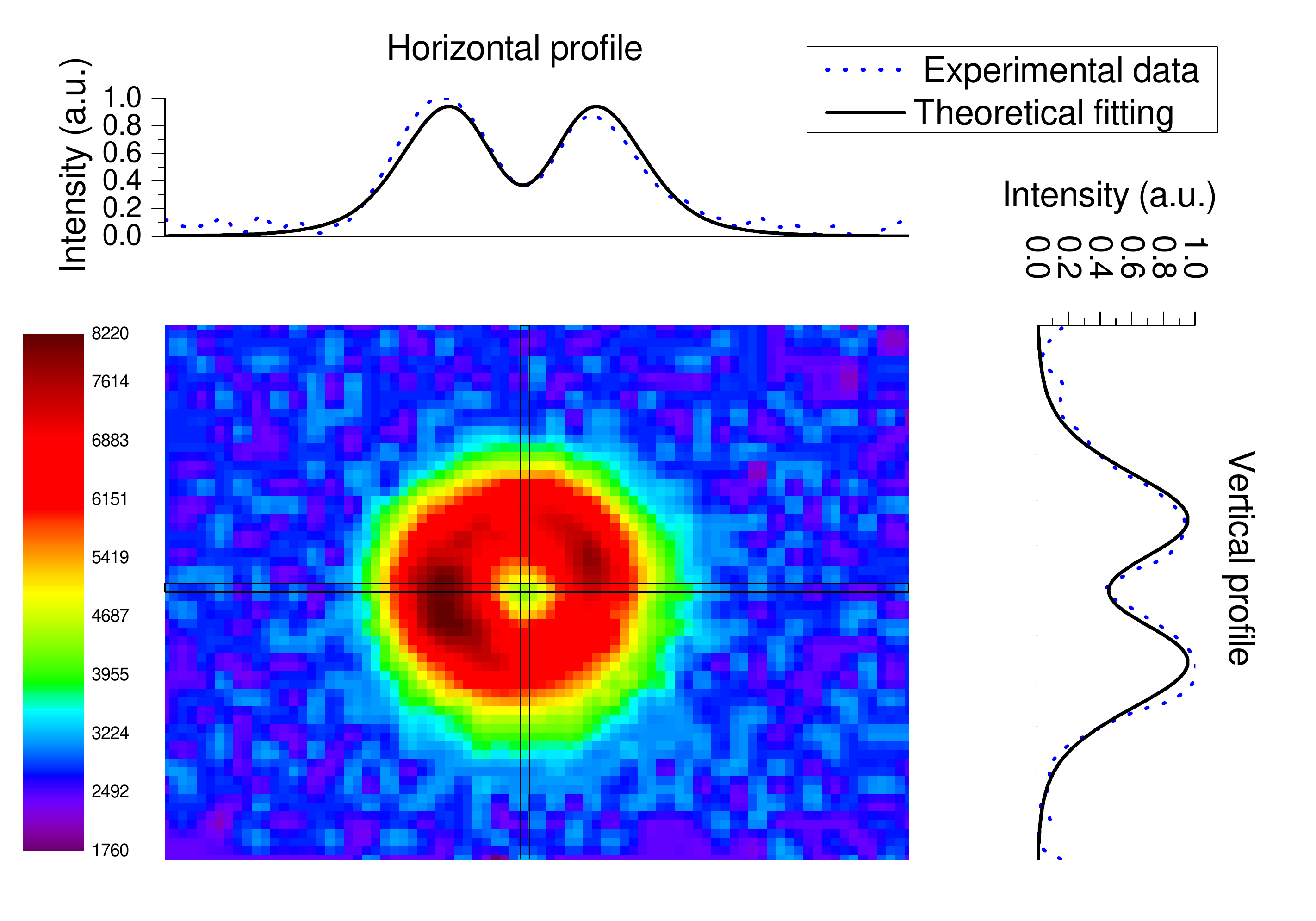}
     \caption{ \label{fig:coolingbeamcoupling} Intensity of the light collected by a CCD camera at one output of the nanofibre when only the MOT beams are on. Horizontal and vertical profiles are fitted using a combination of  LG$_{01}$ and the fundamental mode.}
\end{figure}

As a first test, in the absence of  cold atoms and  probe light through the FM-ONF, an analysis of MOT beam coupling to the nanofibre was conducted. The mode profile of the coupled light was analysed at the output of one of the FM-ONF  pigtails by focussing it onto a CCD camera (replacing the SPCM by a CCD camera, in Fig. \ref{fig:Schematic of the experimental setup} ). The observed profile, as shown in Fig. \ref{fig:coolingbeamcoupling}, was doughnut shaped, but not completely dark at the centre.  A fit of the observed intensity profiles using a combination of the LG$_{00}$ and LG$_{01}$ modes (where the LG$_{00}$ mode corresponds to the LP$_{01}$ fibre mode and the LG$_{01}$ mode to the LP$_{11}$ fibre mode) revealed that $\sim$55\% of the light coming out from the nanofibre pigtail was in  LG$_{01}$. Taking the transmission difference for the fundamental and the higher modes through the FM-ONF into consideration, we estimate that $\sim$67\% of the light from the MOT beams that coupled into the nanofibre excited higher order modes.

 \subsection{Fluorescence measurements}

 \begin{figure}[htbp]
\centering \includegraphics[width= 14 cm]{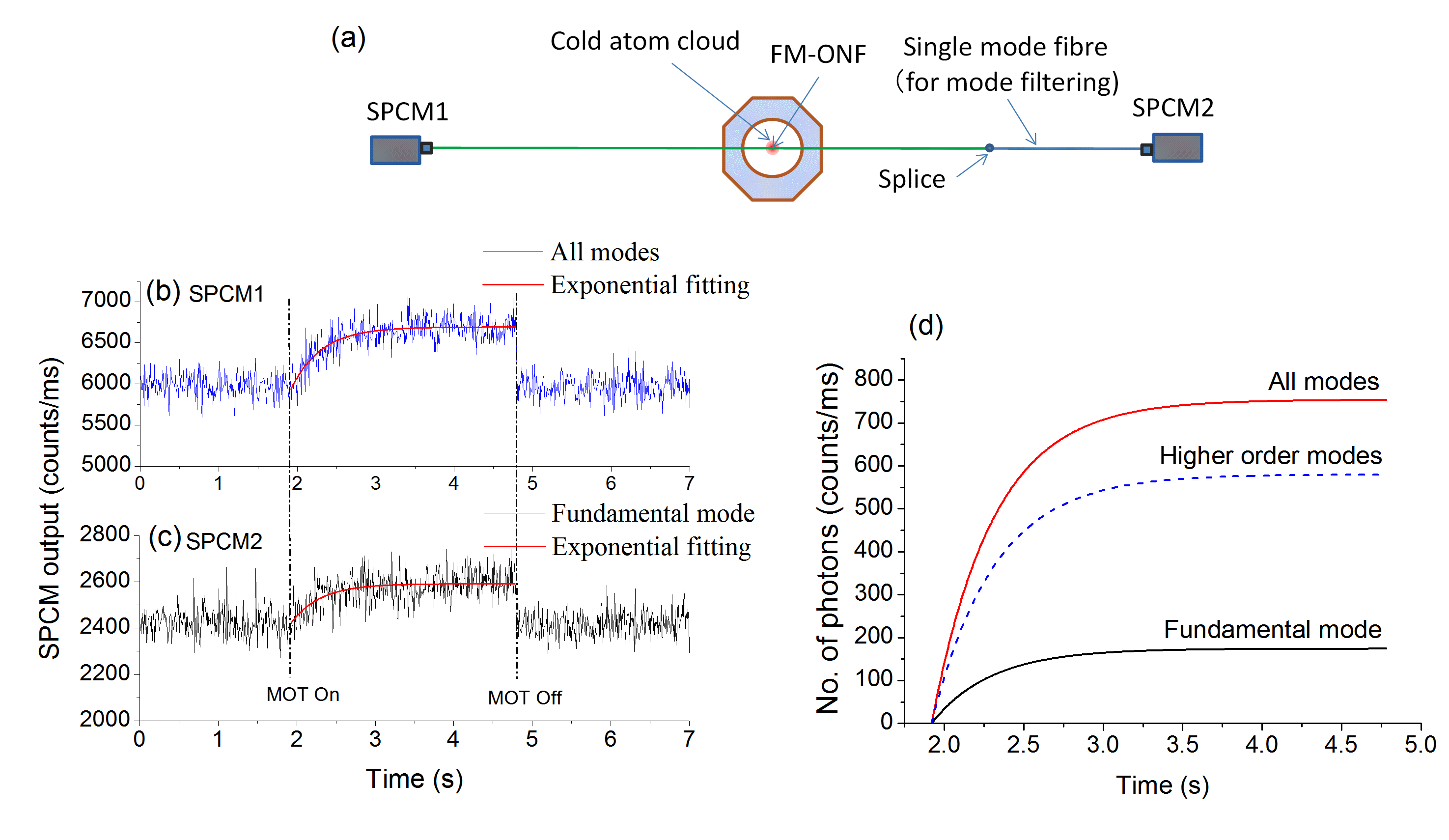}
     \caption{ \label{fig:fluorescence} (a) Schematic of the experimental setup for the mode filtering experiment; (b) Cold atom cloud loading curve when light is coupled to the guided modes of the FM-ONF; (c) the same as (b) only for the fundamental mode. When the magnetic field is switched on, the cloud starts loading and the photon count increases exponentially until it reaches a steady state when the cloud is fully loaded.  After reaching this level, the MOT magnetic field is switched off.  All the atoms leave the trap and the recorded photon count rate is purely due to the MOT beams coupling into the nanofibre; (d) Comparison of fluorescence counts based on the exponential fits to the loading curves in (b) and (c). The curve for the higher order modes (blue, dashed) is generated by subtracting the fundamental mode contribution from all modes. }.
\end{figure}

Next, we looked at the fluorescence coupling into the FM-ONF from resonantly-excited atoms. The cold atom cloud was formed around the waist of the FM-ONF and both the output pigtails were connected to single photon counting modules (SPCMs). In this condition, light coupled into the nanofibre had a contribution from (i) the MOT beams and (ii) the atomic fluorescence. Photon counts were recorded on both the SPCMs and the signals were found to be equivalent, i.e.  half of the photons coupled in to the nanofibre travelled in each direction.  In order to estimate the contribution of the fundamental mode to the total photon count rate, one output pigtail was spliced to a section of single mode fibre (SMF, 780HP, Thorlabs).  The SMF acted as a filter for any higher order mode propagation (Fig. \ref{fig:fluorescence} (a)) and only fundamental mode guiding survived. Simultaneously, the second SPCM recorded the total number of photons coupled into all the fibre modes, i.e. both the fundamental and higher orders, collectively. The MOT magnetic field was switched on and off at intervals of few seconds (to provide enough time for the cloud to reach steady state) in order to separate the photon count contribution from the atom cloud (as shown in Figs. \ref{fig:fluorescence} (b) and (c)) and that arising from the MOT beams.  By comparing the total photon signal with that obtained for only the fundamental mode, we determined that $\sim$85\% of the atomic fluorescence coupling to the nanofibre was coupled into the higher order modes.  In other words, for every photon coupled into the fundamental mode, approximately six photons coupled into the higher fibre modes.  Note that the different transmissions from the waist to the pigtail output for the fundamental mode (96\%) versus the higher order modes (56\%) were taken into consideration. 

In order to confirm this result, the same test was repeated with a different method for filtering out the higher order modes contribution. A SM-ONF was spliced to the output pigtail of the FM-ONF in place of the mode filtering SMF. The SM-ONF was fabricated from the same few-mode fibre (SM1250G80) as the nanofibre in the chamber. Similar results were obtained (data not shown).

\subsection{Absorption measurements}
Next, we considered atom absorption of the light in the evanescent field to see how it varied depending on whether higher order modes or the fundamental mode were guided by the nanofibre. A fraction of the cooling laser beam was passed through a double-pass AOM so that its detuning could be changed as required. The output of the AOM was passed through a linear polarizer before reflecting from the phase imprinted SLM to generate an LG$_{01}$ probe beam, which was coupled to the FM-ONF to excite the higher order modes.  The computer-controlled SLM was used to switch the free-space probe beam between LG$_{00}$ and LG$_{01}$ depending on need.

\begin{figure}[htbp]
\centering \includegraphics[width=5 cm]{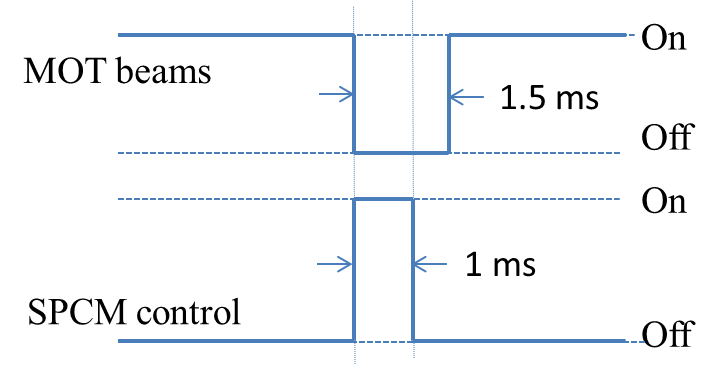}
     \caption{ \label{fig:experiment sequence} Timing sequence for the MOT beams and the SPCM during absorption experiments.  A 5 Hz repetition rate was used.}
\end{figure}

The cold atom cloud was formed around the waist of the nanofibre, ensuring that the atoms in the cloud interacted with the guided light via evanescent field coupling. The MOT beams, i.e. the cooling and  repump beams, and the photon counter were switched on and off using the timing sequence shown in Fig. \ref{fig:experiment sequence} in order to check absorption of the probe beam by the cold atoms. The trapping magnetic field and the probe beam were kept on at all times; however, the photons guided through the nanofibre were only counted during the 1 ms time when the MOT beams were off. A repetition rate of 5 Hz was used for the experiments to ensure that the atom cloud was fully loaded  before proceeding with any measurements. This gave sufficient time for the cloud to return to its  steady state by re-collecting any atoms lost in the 1 ms expansion during the detection phase. The photon counts were collected for 400 runs.

\begin{figure}[htbp]
\centering \includegraphics[width=8 cm]{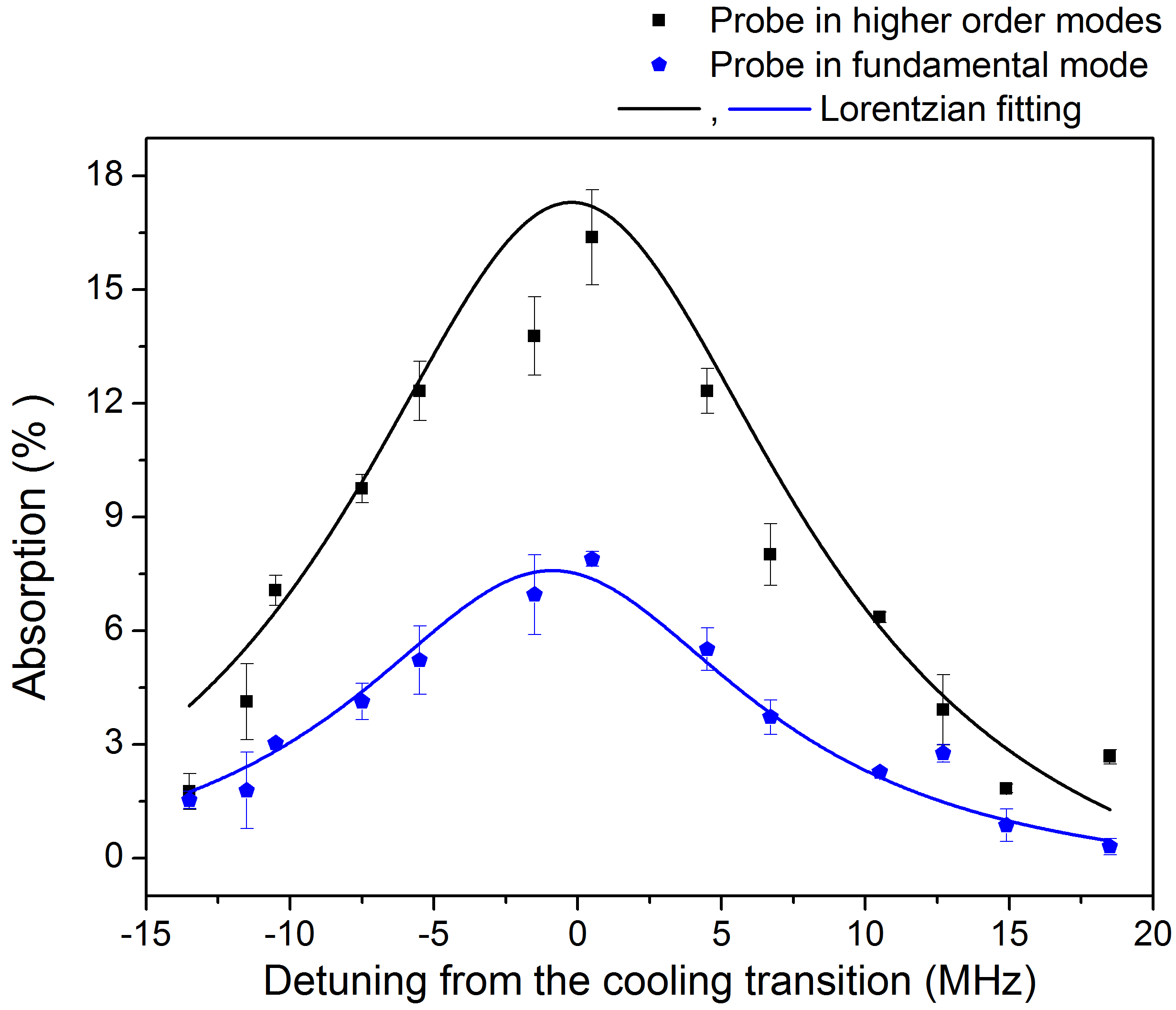}
     \caption{ \label{fig:Absorption spectra} Absorption spectra obtained when either higher order modes (squares) or a fundamental mode (circles) probe beam was coupled into the FM-ONF. Probe power measured at the output end was maintained at $\sim$ 4.3 pW in both  cases by changing the input power.}
\end{figure}

Next, the same collection sequence (Fig. \ref{fig:experiment sequence}) was repeated 400 times in the absence of an atom cloud, so as to determine the background signal. Using these two sets of data, the percentage absorption of the probe was calculated for a particular detuning with respect to the cooling transition of  $^{87}$Rb. The same experiment was repeated using different probe beam detunings in order to obtain the absorption spectrum for the $^{87}$Rb cloud (Fig. \ref{fig:Absorption spectra}). Finally, a similar experiment was performed for a fundamental mode probe in the nanofibre. We used the same power values at the output end of the nanofibre. Again, the percentage absorption signal for the atom cloud was obtained. From Fig. \ref{fig:Absorption spectra} one can see that the absorption appeared more pronounced when the LP$_{11}$ mode group was at the nanofibre waist.  For all cases the probe power was maintained at 4.3 pW at the output end in the cloud-off condition. Taking the transmission difference for the LP$_{00}$ and LP$_{01}$  modes into account, the probe power at the waist of the nanofibre for the higher order mode case was 1.7 times higher than for the fundamental mode.  The absorption percentage for the higher order modes still appeared to be higher (by a factor of $\sim $2). This reflects the fact that more atoms surrounding the waist of the nanofibre interacted with the evanescent field for higher order modes, since they extend further from the fibre surface. The linewidths obtained by Lorentzian fitting to the spectra are similar in both the cases (19 $\pm$ 5 MHz for the higher modes and 17 $\pm$ 2 MHz for the fundamental mode).

\section{Conclusion}
In this work, we have demonstrated, for the first time, the propagation of higher order modes (the LP$_{11}$ mode group) in an optical nanofibre integrated into a magneto-optical trap for neutral atoms. We have studied fluorescence from atoms coupled into the optical nanofibre guided modes. Our preliminary results appear to be in qualitative agreement with earlier theoretical predictions  \cite{Masalov2013, Masalov2014}, but further studies are necessary to fully understand the effects.  In particular, state selection of the atoms should be performed \cite{Kien2014}. We also observed absorption of the higher modes and, for a particular output power level, absorption appeared to be higher than that for  the fundamental mode. The work presented in this paper enables numerous heretofore theoretical atom trapping schemes to finally be realised, such as that relying on modal interference of far detuned higher order modes \cite{Fu2008}. Future work will focus on trapping atoms around the nanofibre using higher order mode combinations and will include consideration of alternative complex mode patterns \cite{ref17}.

\section{Acknowledgments}
This work was supported by the OIST Graduate University and Science Foundation Ireland under Grant No.08/ERA/I1761 through the NanoSci- E+ Transnational Programme, NOIs.


\end{document}